\documentstyle[epsf]{mn}

\newif\ifAMStwofonts

\def\rs{r_{\rm S}}
\def\rhobar{\bar{\rho}}
\def\nbar{\bar{n}}
\def\nbarr{\bar{n}_{\rm R}}
\def\nbarb{\bar{n}_{\rm B}}
\def\nuprime{\nu^{\prime}}
\def\numgal{\langle {\rm N} \rangle}
\def\numgalr{\langle {\rm N} \rangle_{\rm R}}
\def\numgalb{\langle {\rm N} \rangle_{\rm B}}
\def\secgal{\langle {\rm N}({\rm N}-1) \rangle}
\def\secgalr{\langle {\rm N}({\rm N}-1)\rangle_{\rm R}}
\def\secgalb{\langle {\rm N}({\rm N}-1)\rangle_{\rm B}}
\def\Nr{N_{\rm R0}}
\def\Nb{N_{\rm B0}}
\def\Mr{M_{\rm R0}}
\def\Mb{M_{\rm B0}}
\def\Mbs{M_{\rm Bs}}
\def\gammar{\gamma_{\rm R}}
\def\gammab{\gamma_{\rm B}}
\def\rhos{\rho_{\rm S}}
\def\rhor{\rho_{\rm SR}}
\def\rhob{\rho_{\rm SB}}
\def\alphar{\alpha_{\rm R}}
\def\alphab{\alpha_{\rm B}}
\def\yr{y_{\rm R}}
\def\yb{y_{\rm B}}
\def\fr{f_{\rm R}}
\def\fb{f_{\rm B}}
\def\brb{b_{\rm RB}(k)}
\def\brg{b_{\rm RG}(k)}
\def\bbg{b_{\rm BG}(k)}
\def\brbg{b_{\rm RBG}(k)}
\def\Plin{P_{\rm LIN}(k)}
\def\Pgg{P_{\rm GG}(k)}
\def\Prr{P_{\rm RR}(k)}
\def\Pbb{P_{\rm BB}(k)}
\def\Prb{P_{\rm RB}(k)}
\def\Phhgg{P^{hh}_{\rm GG}(k)}
\def\PPgg{P^{P}_{\rm GG}(k)}
\def\Phhrr{P^{hh}_{\rm RR}(k)}
\def\PPrr{P^{P}_{\rm RR}(k)}
\def\Phhbb{P^{hh}_{\rm BB}(k)}
\def\PPbb{P^{P}_{\rm BB}(k)}
\def\Phhrb{P^{hh}_{\rm RB}(k)}
\def\PPrb{P^{P}_{\rm RB}(k)}

\ifoldfss
  \ifCUPmtlplainloaded \else
    \NewTextAlphabet{textbfit} {cmbxti10} {}
    \NewTextAlphabet{textbfss} {cmssbx10} {}
    \NewMathAlphabet{mathbfit} {cmbxti10} {} 
    \NewMathAlphabet{mathbfss} {cmssbx10} {} 
  \fi
  \ifAMStwofonts
    \ifCUPmtlplainloaded \else
      \NewSymbolFont{upmath} {eurm10}
      \NewSymbolFont{AMSa} {msam10}
      \NewMathSymbol{\upi}     {0}{upmath}{19}
      \NewMathSymbol{\umu}     {0}{upmath}{16}
      \NewMathSymbol{\upartial}{0}{upmath}{40}
      \NewMathSymbol{\leqslant}{3}{AMSa}{36}
      \NewMathSymbol{\geqslant}{3}{AMSa}{3E}

    \fi
  \fi
\fi 

\ifnfssone
  \newmathalphabet{\mathit}
  \addtoversion{normal}{\mathit}{cmr}{m}{it}
  \addtoversion{bold}{\mathit}{cmr}{bx}{it}
  \newmathalphabet{\mathbfit} 
  \addtoversion{normal}{\mathbfit}{cmr}{bx}{it}
  \addtoversion{bold}{\mathbfit}{cmr}{bx}{it}
  \newmathalphabet{\mathbfss} 
  \addtoversion{normal}{\mathbfss}{cmss}{bx}{n}
  \addtoversion{bold}{\mathbfss}{cmss}{bx}{n}
  \ifAMStwofonts
    \ifCUPmtlplainloaded \else
      \UseAMStwoboldmath
      \makeatletter
      \new@mathgroup\upmath@group
      \define@mathgroup\mv@normal\upmath@group{eur}{m}{n}
      \define@mathgroup\mv@bold\upmath@group{eur}{b}{n}
      \edef\UPM{\hexnumber\upmath@group}
      \new@mathgroup\amsa@group
      \define@mathgroup\mv@normal\amsa@group{msa}{m}{n}
      \define@mathgroup\mv@bold\amsa@group{msa}{m}{n}
      \edef\AMSa{\hexnumber\amsa@group}
      \makeatother
      \mathchardef\upi="0\UPM19
      \mathchardef\umu="0\UPM16
      \mathchardef\upartial="0\UPM40
      \mathchardef\leqslant="3\AMSa36
      \mathchardef\geqslant="3\AMSa3E
    \fi
  \fi
\fi 

\ifnfsstwo
  \DeclareMathAlphabet{\mathbfit}{OT1}{cmr}{bx}{it}
  \SetMathAlphabet\mathbfit{bold}{OT1}{cmr}{bx}{it}
  \DeclareMathAlphabet{\mathbfss}{OT1}{cmss}{bx}{n}
  \SetMathAlphabet\mathbfss{bold}{OT1}{cmss}{bx}{n}
  \ifAMStwofonts
    \ifCUPmtlplainloaded \else
      \DeclareSymbolFont{UPM}{U}{eur}{m}{n}
      \SetSymbolFont{UPM}{bold}{U}{eur}{b}{n}
      \DeclareSymbolFont{AMSa}{U}{msa}{m}{n}
      \DeclareMathSymbol{\upi}{0}{UPM}{"19}
      \DeclareMathSymbol{\umu}{0}{UPM}{"16}
      \DeclareMathSymbol{\upartial}{0}{UPM}{"40}
      \DeclareMathSymbol{\leqslant}{3}{AMSa}{"36}
      \DeclareMathSymbol{\geqslant}{3}{AMSa}{"3E}
    \fi
  \fi
\fi 

\ifCUPmtlplainloaded \else
  \ifAMStwofonts \else 
    \def\upi{\pi}
    \def\umu{\mu}
    \def\upartial{\partial}
  \fi
\fi

\begin{document}

\title{Modeling Galaxy Clustering by Color}

\author[R.~Scranton]{Ryan Scranton$^{1,2}$ 
\vspace{1mm}\\
$^1$ Department of Astronomy and Astrophysics, University of Chicago, Chicago, 
IL 60637 USA\\
$^2$ NASA/Fermilab Astrophysics Center, P.O. Box 500, Batavia, IL 60510 USA\\
scranton@oddjob.uchicago.edu}

\maketitle 
 
\begin{abstract}

We extend the mass-halo formalism for analytically generating power spectra 
to allow for the different clustering behavior observed in galaxy 
sub-populations.  Although applicable to other separations, we concentrate our 
methods on a simple separation by rest-frame color into ``red'' and ``blue'' 
sub-populations through modifications to the $\numgal(M)$ relations and halo 
distribution functions for each of the sub-populations.  This sort of 
separation is within the capabilities of the current generations of simulations
as well as galaxy surveys, suggesting a potentially powerful observational 
constraint for current and future simulations.  In anticipation of this, we 
demonstrate the sensitivity of the resulting power spectra to the choice of 
model parameters.

\end{abstract}

\bigskip


\section{Introduction}

The difference in clustering between intrinsically red and blue galaxies has
been known since galaxies were first classified into types by Hubble.  More 
recently, the state of cosmological simulations has reached the level where 
galaxy evolution can be modeled semi-analytically (White \& Rees, 1978; White 
\& Frenk, 1991) to produce simulated catalogs with realistic distributions of 
galaxy colors and types (Kauffmann et al., 1999; Somerville \& Primack, 1999;
Benson et al., 2000;).  Within the framework of these types of simulations, 
the power spectra and bias of ``red'' and ``blue'' galaxies have been measured 
and found to be in relatively good agreement with previous data measurements.  
In this paper we seek to derive an analytic method for generating power 
spectra and biases for these red and blue galaxies from within the mass-halo 
model.

The first papers developing the revisions of the original mass-halo formalism 
for dark matter and galaxies focused on real-space correlation functions 
(Sheth \& Jain, 1997; Jing et al., 1998; Peacock \& Smith 2000).  Recently, 
these treatments have been extended for the calculation of power spectra 
(Seljak, 2000; Scoccimarro et al., 2000; Ma \& Fry, 2000).  Although the 
results generated by these models closely matched the observed power law 
behavior for galaxy power spectra (Hamilton et al., 2001), they did not 
address the observed different clustering within sub-populations.  We build on 
the results from these earlier power spectra treatments to produce physically 
motivated models for power spectra and biases for galaxy sub-populations.  In 
an effort to keep our model as general as possible, we will refrain from 
specifying precisely what constitutes a ``red'' or ``blue''galaxy; our 
approach should work equally well for galaxies separated by type or any other 
observable with different clustering properties.

In \S\ref{sec:calc}, we review the formalism for calculating galaxy power
spectra.  With this laid out, \S\ref{sec:mod} details the modifications to 
the halo profiles and the number-mass relation for galaxies populating the 
halos.  Additionally, we work through the modifications to the formulae from 
\S\ref{sec:calc} necessary to calculate the sub-population power spectra and
cross-power spectrum, as well as the associated relative biases. Finally, in 
\S\ref{sec:variation}, we explore the power spectrum space spanned by the
new parameters added to the standard model in \S\ref{sec:mod}.

\section{Calculating Power Spectra}\label{sec:calc}

Following the treatment in Seljak, we have four essential ingredients in our
galaxy power spectrum: the halo profile, the halo mass function, the
halo biasing function, and the galaxy number function.  Once these four 
components are determined, either from observations or simulations, we can 
fold them together to produce the predicted power spectrum for that particular
model. 

We begin by generating a halo profile, $\rho(r)$, which is parametrized along 
the lines of the profile derived by Navarro, Frenk and White (1996; NFW,
hereafter),
\begin{equation}
\rho(r) = \frac{\rhos}
{ \left ( r/ \rs \right )^{-\alpha} \left ( 1 + r/ \rs \right )^{(3+\alpha)}}
\label{eq:NFW}
\end{equation}
where $\rs$ is the universal scale radius and $\rhos = \rho(\rs)$.  In most
halo-model treatments, $\rs$ is replaced by a concentration, $c\equiv r_v/\rs$,
where $r_v$ is the virial radius defining the region where the fractional 
overdensity of the halo $\delta_v$ is approximately 200 and $c$ is a weak
function of halo mass ($c\equiv c_0(M/M_*)^\beta$, where 
$c_0 \sim {\cal O}(10)$ and $\beta \sim -{\cal O}(10^{-1})$).  The 
traditional NFW profile gives $\alpha = -1$, while the Moore profile 
(Moore et al. 1998) has $\alpha = -3/2$.  We will use $\alpha = -1.3$ for the 
calculations in this paper, but the general results are largely insensitive to
the choice of $\alpha$.  Bullock (2001) gives $c_0 = 9$ for a pure NFW profile;
using Peacock \& Smith's relation, $c_0 \approx 4.5$ for a Moore profile.  
Since we are using an intermediate value of $\alpha$, we choose $c_0 = 6$ and 
$\beta = -0.15$ for all the calculations in this paper.  In principle, one
can also consider scatter in the concentration at a given mass (Scoccimarro, 
et al., 2001), leading to an integral over the distribution of $c$, but the 
magnitude of this effect is small enough that it can be safely ignored.

Rewriting Equation~\ref{eq:NFW} in terms of the concentration and the mass, we 
get
\begin{equation}
\rho(r,M) = \frac{\rhos}{\left ( rc/r_v \right)^{-\alpha} \left (1 + rc/r_v 
\right)^{3+\alpha} },
\end{equation}
where 
\begin{eqnarray}
r_v^3 &=& \frac{3M}{4 \pi \delta_v \rhobar}, \\
\rhos &=& \frac{\delta_v \rhobar c^3(M)}{3} 
\left [\int_0^{c(M)} d\chi \frac{\chi^{2+\alpha}}{(1+\chi)^{3+\alpha}} 
\right]^{-1},
\end{eqnarray}
$\bar{\rho}$ is the mean matter density and $M$ is the mass of the halo.  
Since we will be working in wavenumber space when we generate the power 
spectrum, we actually need to consider the Fourier transform of the halo 
profile,
\begin{equation}
y(k,M) = \frac{1}{M} \int_{0}^{r_v} 4 \pi r^2 \rho(r,M) \frac{\sin (kr)}{kr} 
dr, 
\label{eq:rho_tilde}
\end{equation}
where we have normalized over mass so that $y(0,M) = 1$ and $y(k > 0, M) < 1$.
Note that this implies that $\rho(r > r_v) = 0$, truncating the mass
integration at the virial radius. 

With this in hand, we can move on to the next component of the halo model,
the halo mass function, ($dn/dM$).  Traditionally, this mass function is 
expressed in terms of a function $f(\nu)$,
\begin{equation}
\frac{dn}{dM} dM = \frac{\rhobar}{M} f(\nu) d\nu,
\end{equation}
where $\nu$ relates the minimum spherical over-density that has collapsed at a
given redshift ($\delta_c(z)$, $\delta_c(0) = 1.676$ for an $\Omega_m = 0.3$, 
$\Omega_\Lambda = 0.7$ cosmology) and the rms spherical fluctuations 
containing mass $M$ ($\sigma(M,z)$) as
\begin{equation}
\nu \equiv \left (\frac{\delta_c(z)}{\sigma(M)} \right )^2.
\end{equation}
We define $M_*$ as the mass corresponding to $\nu = 1$.  The functional form
for $f(\nu)$ is traditionally given by the Press-Schechter function (1974).  
This form tends to over-predict the number of halos below $M_*$, so we use the 
form found from simulations by Sheth and Torman (1997),
\begin{equation}
\nu f(\nu) \sim (1 + {\nuprime}^{-p}){\nuprime}^{1/2} e^{-{\nuprime}/2},
\end{equation}
where $\nuprime = a\nu$, $a = 0.707$ and $p = 0.3$.  This relation is
normalized by requiring that 
\begin{equation}
\frac{1}{\rhobar} \int_{0}^{\infty} \frac{dn}{dM} M dM = \int f(\nu) d\nu = 1,
\label{eq:st_norm}
\end{equation}
for the dark matter distribution. 

On nonlinear scales, we expect the halos to cluster more strongly than the 
mass, and vice versa for linear scales (Mo \& White, 1996).  This means we 
need to positively bias the clustering of the high mass halos relative to the 
low mass halos.  We can generate this sort of halo biasing scheme for the ST 
mass function using
\begin{equation}
b(\nu) = 1 + \frac{\nuprime - 1}{\delta_c} + 
\frac{2p}{\delta_c(1 + {\nuprime}^p)}.
\label{eq:halo_bias}
\end{equation}
In order for the eventual power spectrum to reduce to a linear power spectrum
on large scales, we need to impose the further constraint that 
\begin{equation}
\int_{0}^{\infty} f(\nu) b(\nu) d\nu = 1,
\label{eq:bias-norm}
\end{equation}
requiring that the biased halos with mass greater than $M_*$ be balanced out
by anti-biased halos with mass less than $M_*$.  This integral is satisfied 
automatically if we use Equation~\ref{eq:halo_bias} and have properly 
normalized $f(\nu)$.

Using just these three components, we can generate the power spectrum for the
dark matter.  However, in order to predict the galaxy power spectrum, we need
to know how many galaxies are in a given halo (under the assumption that 
the distribution of galaxies in the halo follows the halo profile).  Currently
no theory completely informs the formation of galaxies given a halo mass, but
the traditional form of the $\numgal(M)$ relation (Jing et al., 1998; 
Kauffmann et al., 1999; Benson et al., 2000; White et al., 2001) has the 
galaxies populating the halo as a simple power law, 
$\numgal(M) \sim (M/M_0)^\gamma$, where $M_0$ sets the unit mass scale and 
$\gamma < 1$.  Additionally, one can put in constraints on the minimum mass to 
form a galaxy and other modifications.  For the purposes of the formalism for 
calculating the power spectrum, however, we can put aside the question of 
precisely what this function looks like.  The inclusion of galaxies does 
change the normalization of 
Equation~\ref{eq:st_norm} to
\begin{equation}
\int_{0}^{\infty} \frac{\numgal}{M(\nu)} f(\nu) d\nu = \frac{\nbar}{\rhobar},
\label{eq:f-norm}
\end{equation}
where $\nbar$ is the mean number of galaxies.

On large scales, the power spectrum is dominated by correlations between 
galaxies in separate halos.  We need to convolve the halo profile with the
mass function to account for the fact that halos are not pointlike objects.
Since we are in Fourier space, we can perform the convolution using simple
multiplication.  The halo-halo power ($\Phhgg$) is then simply,
\begin{equation}
\Phhgg = \Plin \left [ \frac{\rhobar}{\nbar} \int_{0}^{\infty} f(\nu) 
\frac{\numgal}{M(\nu)} b(\nu) y(k,M) d\nu \right ]^2,
\label{eq:halo-halo}
\end{equation}
where $\Plin$ is the linear dark matter power spectrum.  In the small $k$
limit, this reduces to a simple linear bias ($\langle b \rangle$),
\begin{equation}
\langle b \rangle = \frac{\rhobar}{\nbar} \int_{0}^{\infty} f(\nu) 
\frac{\numgal}{M(\nu)} b(\nu) d\nu.
\end{equation}

For small scales, the dominant contribution to the power spectrum comes from 
correlations between galaxies within the same halo.  This single halo term is 
independent of $k$ at larger scales, giving it a Poisson-like behavior.  In 
order to account for the fact that a single galaxy within a halo does not
correlate with itself, we use the second moment of the galaxy number relation,
($\secgal$) to calculate the Poisson power ($\PPgg$),
\begin{equation}
\PPgg = \frac{\rhobar}{(2\pi)^3 \nbar^2} \int_{0}^{\infty} f(\nu) 
\frac{\secgal}{M} |y(k,M)|^\zeta d\nu.
\label{eq:poisson}
\end{equation}
Seljak takes $\zeta = 2$ for $\secgal > 1$ and $\zeta = 1$ for 
$\secgal < 1$; this is done to account for the galaxy at the center of the 
halo in the limit of small number of galaxies.  In the limit of large numbers 
of galaxies, $\secgal$ approaches $\numgal^2$, but in the small number limit, 
$\secgal$ can be approximated by a binomial distribution.  This can be 
implemented by using the fits given in Scoccimarro et al. (2000), letting
\begin{equation}
\secgal = \alpha_M^2 \numgal^2,
\end{equation}
where $\alpha_M = 1$ for $M > 10^{13}h^{-1}M_{\sun}$ and 
$\alpha_M = \log(\sqrt{Mh/10^{11}M_{\sun}})$ for $M < 10^{13}h^{-1}M_{\sun}$.  
Clearly, this will not be exact for an arbitrary $\numgal(M)$, but it should 
be close enough for our purposes.  Adding $\Phhgg$ and $\PPgg$, we recover the 
galaxy power spectrum at all wavenumbers, $\Pgg$.  

\section{Generating Red \& Blue Power Spectra}\label{sec:mod}

	Within the framework presented above, there are a number of parameters 
which might be modified to generate models of different power spectra for red 
and blue galaxies.  One could modify the concentration index for each galaxy 
population, change the halo biasing relation, etc.  In this paper, we focus on 
two modifications to the standard treatment: a modification of the $\numgal$ 
relations and the halo profiles for each of the galaxy types.  

The physical motivation in both of these cases is clear.  Semi-analytic models
for galaxy formation indicate that the primary determinant of galaxy color is
the epoch of initial gas cooling to form the initial stellar population.  
Red galaxies tend to form earlier, appearing in the deepest over-densities, 
while the current blue galaxies form later when gas in the shallower 
potentials and outskirts of the larger potentials has cooled.  Given this 
difference in development, the prospect that the efficiency of galactic 
formation (and hence number of galaxies produced within a halo of a given 
local halo mass) would vary for the two epochs is a reasonable conclusion.  
Likewise, for the different halo profiles, we know from observations as well 
as simulations that red galaxies tend to populate the centers of galaxy 
clusters and filaments, while blue galaxies are more numerous at the fringes 
of structure and in the field.  Changing the distribution function for the 
different colored galaxies within a halo to reflect these observations is an 
obvious step.  

\subsection{Modifying $\numgal(M)$}

The simplest model we can adopt for the galaxy number relations for the red 
and blue galaxies ($\numgalr$ and $\numgalb$, respectively) would be the 
simple power laws alluded to earlier,
\begin{eqnarray}
\numgalr(M) = \Nr \left ( \frac{M}{\Mr} \right )^{\gammar} \\
\numgalb(M) = \Nb \left ( \frac{M}{\Mb} \right )^{\gammab}. \nonumber
\end{eqnarray}
In practice, this form actually over-determines the functional form of the 
power laws; since we know the functions pass through unity at some mass scale,
we can set $\Nr = \Nb = 1$ and determine the relative contribution of red and 
blue galaxies by selecting $\Mr$ and $\Mb$ appropriately.  

Such a model does reasonably well, but the GIF simulations (Kauffmann, et al., 
1999) point to an extra abundance of blue galaxies at small halo masses 
($M \sim 10^{12}h^{-1}M_{\sun}$).  Sheth et al. (2001) include this effect by 
the addition of a Gaussian term to the $\numgalb$ relation,
\begin{equation}
\numgalb(M) = \left (\frac{M}{\Mb} \right )^{\gammab} + 
Ae^{-4(\log(M) - \Mbs)^2},
\end{equation}
where $A$ is ${\cal O}(1/2)$ and $\Mbs$ is the logarithm of the mass 
corresponding to the peak in the Gaussian component.  All told, this gives us 
six tunable parameters for the $\numgal$ relations.  For the purposes of our 
power spectrum calculations, we will add an additional lower mass cut-off for 
the $\numgalr$ relations at $10^{11}h^{-1}M_{\sun}$; the resulting power 
spectra are not terribly sensitive to the precise value of this cut-off.

\subsection{Modifying Halo Profiles}

In modifying the distribution functions for red and blue galaxies, we have a
number of constraints.  First, we require that the sum of the matter associated
with red galaxies and that with blue galaxies match the total distribution of
matter at all halo radii.  This is not to say that all the matter ends up in 
galaxies; rather, it merely requires that the combined distribution of red and 
blue galaxies in the halo match the total galaxy distribution.  Although this
would appear to suppress natural correlations between red and blue galaxies,
Sheth \& Lemson (1999) show that this sort of clustering by conservation of 
number works reasonably well.

There are any number of profiles we might choose to consider for the red and 
blue sub-profiles.  In principle, the shapes of these profiles could be found 
from analyzing the results of simulations, but for the purposes of this 
exercise we will forgo that complication.  Instead, we will restrict ourselves 
to profiles of a similar form as that given in Equation~\ref{eq:NFW} with 
different values of $\rhos$ and $\alpha$ for the red and blue sub-populations. 
Of course, the sum of two profiles with differing values of $\alpha$ do not 
quite match a profile with a third value of $\alpha$, but we can come 
reasonably close with clever choices of $\alphar$ and $\alphab$ for the red 
and blue populations, respectively.

In order to find the proper values for our extended halo parameter set, we 
need only know the relative abundance of red and blue galaxies at two points 
along the radial profile.  At large radii relative to the virial radius, all 
of the profiles go as $\rho \sim r^-3$, meaning that the ratio of the number 
of blue galaxies to red galaxies ($\eta$) in this regime gives us the ratio of 
$\rhob/\rhor$ immediately, once we have transformed from number of galaxies to 
mass assigned to galaxies (see below).  We consider the halos truncated at 
these large radii when we calculate the power spectra, but this allows us to 
set the relative normalization of $\rhor$ and $\rhob$ under the assumption 
that $\eta$ does not vary much beyond $r_v$.

With this ratio determined, the choice of the second radial point 
is somewhat arbitrary, but clearly we would prefer this point to be at very 
small radius where the profiles go as $\rho \sim r^\alpha$.  For all of the 
calculations, we set the inner radius such that $r_i c/ r_v = 0.1$.  This 
demands a resolution smaller than most simulations can provide at the moment, 
but does guarantee that we are well into the $\rho \sim r^\alpha$ regime.  

For a given ratio ($\mu$) of red to blue galaxies at small radius, the 
difference in $\alpha$ for the red and blue profiles is 
\begin{equation}
\Delta \alpha \equiv \alphab - \alphar = 
\frac{\log(\mu \eta)}{\log(1 + r_i c/r_v) - \log(r_i c/ r_v) }, 
\label{eq:alpha}
\end{equation} 
where we have assumed that the mass follows the number of galaxies linearly.
However, to properly conserve the mass in the halo, we need to transform $\mu$
and $\eta$ into mass-space using the $\numgal$ functions for the red and blue 
galaxies to yield $\mu^\prime$ and $\eta^\prime$, respectively.  In the 
absence of the Gaussian term in the $\numgalb$ relation, these transform as
\begin{eqnarray}
\label{eq:eta_prime_simple} 
\eta^\prime &=& \eta^{1/\gammab}\frac{\Mb}{\Mr} \\ 
\mu^\prime &=& \mu^{1/\gammar}\frac{\Mr}{\Mb}. \nonumber
\end{eqnarray}
If we wish to include the Gaussian term, a bit of algebra gives the slightly
more complex form of the transformation:
\begin{eqnarray}
\label{eq:eta_prime} 
\eta^\prime &=& \left[ \eta - A e^{-4(\log(\Mr)-\Mbs)^2} \right] ^{1/\gammab}
\frac{\Mb}{\Mr} \\ 
\mu^\prime &=& 
\left [ \mu \left (1 + Ae^{-4(\log(\Mb)-\Mbs)^2} \right) \right ]^{1/\gammar}
\frac{\Mr}{\Mb}. \nonumber
\end{eqnarray}
As one might expect, the limit that $\Mr, \Mb \gg 10^{\Mbs}$, 
Equation~\ref{eq:eta_prime} effectively reduces to 
Equation~\ref{eq:eta_prime_simple}.  Once we have made this transformation, we 
then replace $\mu$ and $\eta$ in Equation~\ref{eq:alpha} with $\mu^\prime$
and $\eta^\prime$.  With this relation between $\alphar$ and $\alphab$ in 
hand, we can perform a simple search over values of $\alphar$ to find the 
sub-profiles that combine to closest match an overall profile with a 
given value of $\alpha$.  Since we know that $\Delta \alpha$ must be positive,
this relation guarantees a flatter distribution of blue galaxies in the 
center of halos relative to red galaxies, something which would be much more 
difficult to accomplish had we varied $c$ for the differ galaxy populations.

\begin{figure}
\begin{center}
\epsfverbosetrue
\epsfysize=2.54in \epsfbox{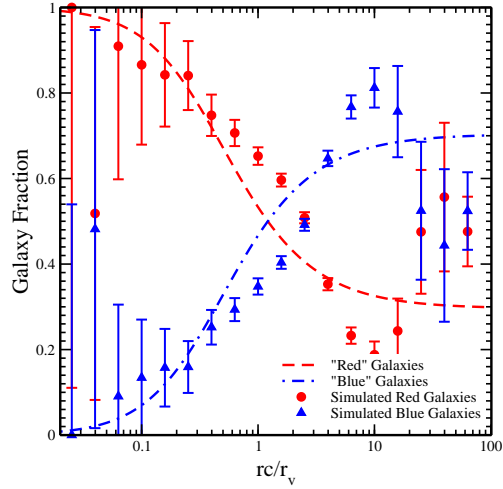}
\caption{Fraction of ``red'' and ``blue'' galaxies as a function of $rc /r_v$
compared to the distribution of galaxies from the GIF simulations.  For the
simulated data, red galaxies were taken to have $g^\prime - r^\prime < 0.6$ 
in the rest frame for SDSS magnitudes (Fukugita, et al. 1996).}
\label{fig:gal_dist}
\end{center}
\end{figure}

Using $\rhor$, $\rhob$, $\alphar$ and $\alphab$, 
where
\begin{eqnarray}
\label{eq:rho_prime} 
\rhor &=& \frac{1}{\eta^\prime + 1} \rhos \\
\rhob &=& \frac{\eta^\prime}{\eta^\prime + 1} \rhos, \nonumber
\end{eqnarray}
we can recalculate $y(k,M)$ along the lines of Equation~\ref{eq:rho_tilde}, 
making $\yr$ and $\yb$.  We can also recover the distribution of galaxies by
color/type within a given halo similar to that seen in simulations 
(cf. Diaferio et al., 1999), as seen in Figure~\ref{fig:gal_dist}.  Clearly,
the agreement is not perfect, but it is close enough that we can proceed with 
confidence that the basic approach is reasonable.  These diagrams inform our 
fiducial choices of $\mu$ and $\eta$ such that $\mu \sim {\cal O}(10)$ and 
$\eta \sim {\cal O}(\sqrt(10))$; for our models we take $\mu = 10$ and 
$\eta = 4$.

\subsection{Calculating Subpopulation Power Spectra and Biases}

Plugging all this into the standard formalism, we can calculate the power
spectra for these galaxy populations, as well as their cross-correlation, in
the same manner as described in Seljak.  In order to calculate the power 
spectra properly, we need to re-normalize $f(\nu)$ for each sub-population 
using Equation~\ref{eq:f-norm} to account for the differences in $\numgal$ and 
$\nbar$:
\begin{eqnarray}
\int_{0}^{\infty} \frac{\numgalr}{M(\nu)} 
\fr(\nu)d\nu &=& \frac{\nbarr}{\rhobar} \\ 
\int_{0}^{\infty} \frac{\numgalb}{M(\nu)} 
\fb(\nu)d\nu &=& \frac{\nbarb}{\rhobar} \nonumber
\end{eqnarray}
Once this has been done, we can insert the above (along with the 
color-dependent halo profiles) into Equations~\ref{eq:halo-halo} and 
\ref{eq:poisson} to generate the power spectra for red and blue galaxies:
\begin{eqnarray}
\frac{\Phhrr}{\Plin} = 
\left [ \frac{\rhobar}{\nbarr} \int_{0}^{\infty} \fr(\nu) 
\frac{\numgalr}{M(\nu)} b(\nu) \yr(k,M) d\nu \right ]^2 \\
\PPrr  = {\rhobar} \int_{0}^{\infty} \fr(\nu) 
\frac{\secgalr}{(2\pi)^3 \nbarr^2 M(\nu)} |\yr(k,M)|^\zeta d\nu, \nonumber \\
\frac{\Phhbb}{\Plin} = 
\left [ \frac{\rhobar}{\nbarb} \int_{0}^{\infty} \fb(\nu) 
\frac{\numgalb}{M(\nu)} b(\nu) \yb(k,M) d\nu \right ]^2 \nonumber \\
\PPbb = \rhobar  \int_{0}^{\infty} \fb(\nu) 
\frac{\secgalb}{(2\pi)^3 \nbarb^2 M(\nu)} |\yb(k,M)|^\zeta d\nu. \nonumber
\end{eqnarray}
As before, we generate the total power spectra ($\Prr$ and $\Pbb$) by taking
the sum of these parts,
\begin{eqnarray}
\Prr &=& \Phhrr + \PPrr \\
\Pbb &=& \Phhbb + \PPbb. \nonumber
\end{eqnarray}

To generate the cross-power spectrum, we take one factor from each of the 
different power spectrum terms.  This makes the halo-halo term
\begin{eqnarray}
\frac{\Phhrb}{\Plin} &=& 
\left [\frac{\rhobar}{\nbarr} \int_{0}^{\infty} \fr(\nu) 
\frac{\numgalr}{M(\nu)} b(\nu) \yr(k,M) d\nu \right ] \times \nonumber \\
 & & \left [\frac{\rhobar}{\nbarb} \int_{0}^{\infty} \fb(\nu) 
\frac{\numgalb}{M(\nu)} b(\nu) \yb(k,M) d\nu \right ]. 
\end{eqnarray}
For the Poisson term, we simply replace the second moment of the galaxy number
relations with the product of $\numgalr$ and $\numgalb$,
\begin{eqnarray}
\PPrb = \rhobar
\int_{0}^{\infty} \bar{f}(\nu) \frac{\numgalr \numgalb}{(2\pi)^3 \nbarr \nbar 
M(\nu)} |\bar{y}(k,M)|^\zeta d\nu,
\end{eqnarray}
where $\bar{f}(\nu)$ and $\bar{y}(k,M)$ are the geometric means of the 
red and blue values.

\begin{figure}
\begin{center}
\epsfysize=2.45in \epsffile{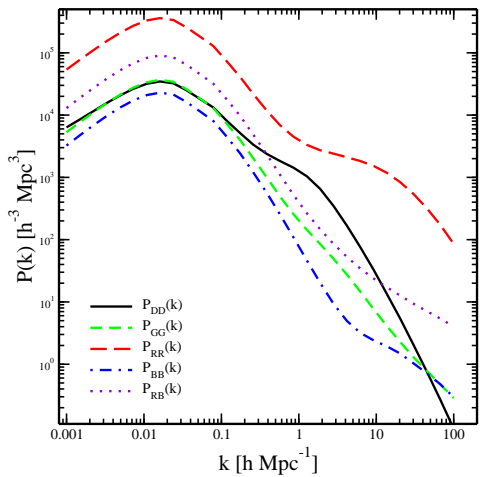}
\caption{Power spectra for red ($\Prr$) and blue ($\Pbb$) galaxies and
their cross-power spectrum ($\Prb$) compared to the dark matter 
($P_{\rm DD}(k)$) and galaxy ($\Pgg$) power spectra.}
\label{fig:power_spectra}
\end{center}
\end{figure}

The results of such a calculation are shown in Figure~\ref{fig:power_spectra}. 
For our fiducial model, we choose the set of input parameters in a 
$\Lambda {\rm CDM}$ model given in Table~\ref{tab:fiducial}.  As we would 
generally expect, the red galaxies show a stronger biasing than either the 
total sample or the blue sample, as well as tracing the shape of the dark 
matter power spectrum ($P_{\rm DD}(k)$) more closely.  The blue galaxies are 
anti-biased relative to the normal galaxy power spectrum, and demonstrate a
slightly steeper slope.  Additionally, the blue galaxies demonstrate a sharp 
break from a power law at small scales.  This effect is due solely to the 
$\numgal(M)$ relation for the blue galaxies and not the halo profiles; the 
larger number of blue galaxies in smaller, less massive halos sets in at this 
scale, driving the power up.  Remarkably, however, the galaxy populations that 
generate these power spectra combine to produce a total galaxy power spectrum 
with simple power law behavior.  The exact comparison of these predicted power 
spectra to those from simulations we leave as a detail for future work; for
now we are more interested in the flexibility of the model than precise values 
for parameters. 

\begin{table}
\begin{center}
\caption{Fiducial model parameters for power spectra in 
Figure~\ref{fig:power_spectra}}
\label{tab:fiducial}
\begin{tabular}{@{}lcc}
Description & Parameter & Value \\
\hline
Red Unit Mass Scale & $\Mr$ & $3\times 10^{12}h^{-1}M_{\sun}$ \\  
Blue Unit Mass Scale & $\Mb$ & $7 \times 10^{13}h^{-1}M_{\sun}$ \\ 
Red Mass Scaling Index & $\gammar$ & 0.9 \\
Blue Mass Scaling Index & $\gammab$ & 0.7 \\  
Gaussian Normalization & $A$ & 0.5 \\  
Gaussian Mass Scale & $\Mbs$ & 11.75  \\
Outer Galaxy Ratio & $\eta$ & 4 \\  
Inner Galaxy Ratio & $\mu$ & 10 \\  
\hline 
\end{tabular}
\end{center}
\end{table}

With this machinery in place, we can calculate the relative bias between the
various power spectra:
\begin{eqnarray}
\brb^2 = \frac{\Prr}{\Pbb} &;& \brg^2 = \frac{\Prr}{\Pgg} \\
\bbg^2 = \frac{\Pbb}{\Pgg} &;& \brbg^2 = \frac{\Prb}{\Pgg}. \nonumber 
\end{eqnarray}
We choose relative biases between the various galaxy power spectra rather than
the absolute biases relative to the dark matter for two reasons.  First, in 
each of the cases in \S\ref{sec:variation} where we vary a parameter in our 
model, at least one of $\Pgg$, $\Prr$ or $\Pbb$ remains roughly constant, so 
we can use that power spectrum as a baseline for seeing how the other one or 
two vary.  Second, while the absolute biases can be measured using galaxy 
magnification bias, the relative biases involve real clustering that can be 
measured over a much wider range of redshift for a given photometric or 
spectroscopic survey (the evolution of these biases over redshift will 
be left for future work).

\begin{figure}
\begin{center}
\epsfysize=2.45in \epsffile{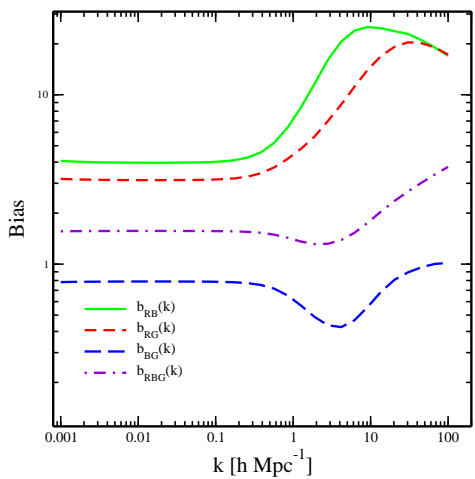}
\caption{Relative biases between red and blue galaxies ($\brb$) and the 
red, blue and cross power spectra and the galaxy power spectrum ($\brg$,
$\bbg$, and $\brbg$, respectively).}
\label{fig:bias}
\end{center}
\end{figure}

As with the results calculated by Seljak, the relative biases shown in 
Figure~\ref{fig:bias} are constant on large scales, whereas on small scales 
there is considerable variation, particularly in the $\brb$ and $\bbg$ 
biases.  As we will see later, the behavior of these biases is a strong 
function of the model input, suggesting that reasonably small error bars on 
the the bias in wavenumber space could act as a powerful constraint on the 
model parameters.

\section{Variations}\label{sec:variation}

\subsection{Inner and Outer Ratios}\label{sec:halo_variation}

\begin{figure}
\begin{center}
\epsfxsize=2.45in \epsffile{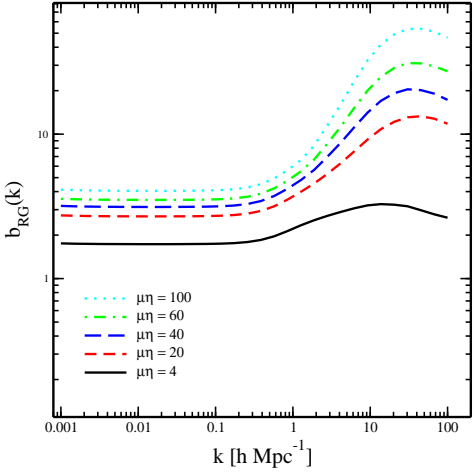}
\caption{$\brg$ for several values of $\mu\eta$ in the limit where 
$\Mb \gg \Mr$.  $\Pgg$ and $\Pbb$ are unaffected by changes in $\mu\eta$ in 
this limit.}
\label{fig:inner_ratio}
\end{center}
\end{figure}

As Equations~\ref{eq:alpha} and \ref{eq:eta_prime} suggest, the effect of our 
choices for $\eta$ and $\mu$ will be at least partially determined by the 
ratio of $\Mr/\Mb$.  In the regime where $\Mb \gg \Mr$ as in the fiducial 
model, the actual values of $\mu$ and $\eta$ are not so important to the 
resulting power spectra as their product.  Likewise, since the profile that 
the blue galaxies will populate is relatively flat in the center of each halo, 
the effects of changing the value of $\mu\eta$ does not significantly change 
the clustering of the blue galaxies; the majority of the mass associated with 
normal NFW profiles is outside the $\rho \sim r^\alpha$ region anyway and, 
since $\alphab > \alpha$ by construction, this will be even more pronounced 
for the blue galaxies.  Thus, we do not expect $\Pbb$ to vary significantly 
with $\mu\eta$ and we have constructed $\alphar$ and $\alphab$ such that 
$\Pgg$ will not vary, so the only sensitivity to $\mu\eta$ we expect to see is 
in $\Prr$ and $\Prb$ and the associated relative biases.  In 
Figure~\ref{fig:inner_ratio}, we show $\brg$ for several different values of 
$\mu\eta$.  There is some change in the shape of the bias, perhaps indicative 
of a more negative $\alphar$ in the high $\mu\eta$ models leading to a greater 
population of the inner regions of the halos with red galaxies.  Indeed, in 
most of these models, the matter associated with red galaxies only exceeds 
that associated with blue galaxies in the very inner regions of the halo.

\begin{figure}
\begin{center}
\epsfxsize=2.45in \epsffile{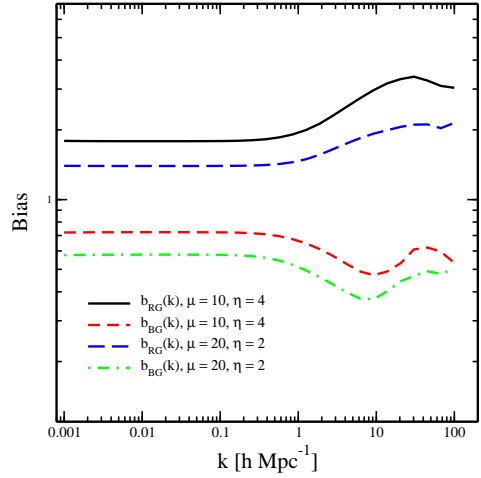}
\caption{$\brg$ and $\bbg$ for equal values of $\mu\eta$ in the 
$\Mr \approx \Mb$ regime.  $\Pgg$ remains constant for all values of $\mu\eta$
by construction.}
\label{fig:mu-eta_degen}
\end{center}
\end{figure}

In a model where $\Mb \approx \Mr$, the degeneracy between $\mu$ and $\eta$ is
broken.  Figure~\ref{fig:mu-eta_degen} shows $\brg$ and $\bbg$ for two models 
where $\Mb = \Mr = 3 \times 10^{12} h^{-1} M_{\sun}$ and $\mu\eta = 40$.
Since Equations~\ref{eq:alpha} and \ref{eq:rho_prime} are no longer dominated 
by the ratio of $\Mr/\Mb$, we can see significant shifts in the biases of both 
red and blue galaxies relative to $\Pgg$.  In this case, we have a more equal 
distribution between the mass associated with red and blue galaxies and less 
extreme relative halo profile normalizations.  Clearly, using future 
measurements to constrain these parameters will require using multiple biases
to minimize these degeneracies.

\subsection{$\numgal$(M) Relations}\label{sec:n_variation}

Unlike modifying the halo profiles, changing the parameters in the 
$\numgal(M)$ relations can have significant effects on the shapes of all the 
power spectra not just $\Prr$ and $\Pbb$.  The general effect of each of the 
modifications is to change the behavior of the Poisson term in the power 
spectra, but the specific effects for each modification show considerable 
and sometimes surprising variations.

\begin{figure}
\begin{center}
\epsfxsize=2.45in \epsffile{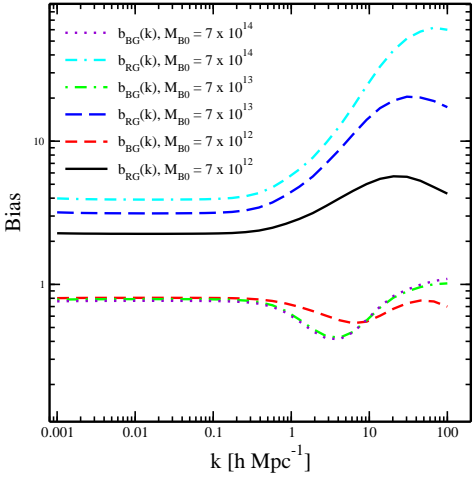}
\caption{$\brg$ and $\bbg$ for several values of $\Mb$.  Masses are given in 
$h^{-1}M_{\sun}$.  $\Pgg$ remains constant over variations in $\Mb$.}
\label{fig:bias_mb}
\end{center}
\end{figure}

We begin with the unit mass scale for the blue galaxies, $\Mb$.  In general,
this parameter does not strongly affect the total galaxy power spectrum; there
is some slight variation ($\sim 5\%$) over the range 
$10^{12}h^{-1}M_{\sun} < \Mb < 10^{14}h^{-1}M_{\sun}$ in the quasi-linear
regime of the power spectrum ($k \approx h Mpc^{-1}$).  However, there is 
significant change in the relative biases of the red and blue galaxies, as
shown in Figure~\ref{fig:bias_mb}.  Additionally, as $\log(\Mb)$ approaches
$\Mbs$, the non-power law behavior of $\Pbb$ is considerably damped out.

\begin{figure}
\begin{center}
\epsfxsize=2.45in \epsffile{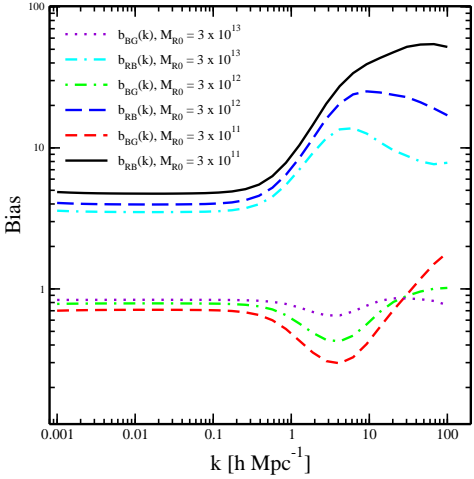}
\caption{$\brb$ and $\bbg$ for several values of $\Mr$.  Masses are given in 
$h^{-1}M_{\sun}$.  $\Pbb$ is constant over this range in $\Mr$.}
\label{fig:bias_mr}
\end{center}
\end{figure}

In contrast to $\Mb$, modifying the values of $\Mr$ leads to large 
variations in the shape and amplitude of $\Pgg$, with lesser amplitude shifts
to $\Prr$ and almost no effect on $\Pbb$ (as we might suspect). 
Figure~\ref{fig:bias_mr} shows $\brb$ and $\bbg$ for a two decade range in 
$\Mr$ values.  As $\Mr$ approaches $\Mb$, the effect of the Gaussian
term in $\numgalb$ on $\Pgg$ increases, leading to a galaxy power spectrum with
a strong break in its power law at large wavenumber.  Conversely, at lower 
$\Mr$, the galaxy power spectrum inflects, leading to a stronger anti-bias in 
$\Pbb$ relative to $\Pgg$ around $k \sim 1$.  Additionally, in this limit we 
can see the effect of the Gaussian component in the transformation from 
$\eta$ to $\eta^\prime$ (Equation~\ref{eq:eta_prime}) changing the effective
value of $\mu\eta$.

Due to the sub-dominant role of $\numgalb$ to that of $\numgalr$ over
most of the mass range, varying $\gammab$ does not significantly change any of
the power spectra.  Changing $\gammar$ alters $\Prr$ and $\Pgg$ slightly, 
generally smoothing out the variations in $\brg$ over $k$ for larger values of
$\gammar$.

\begin{figure}
\begin{center}
\epsfxsize=2.45in \epsffile{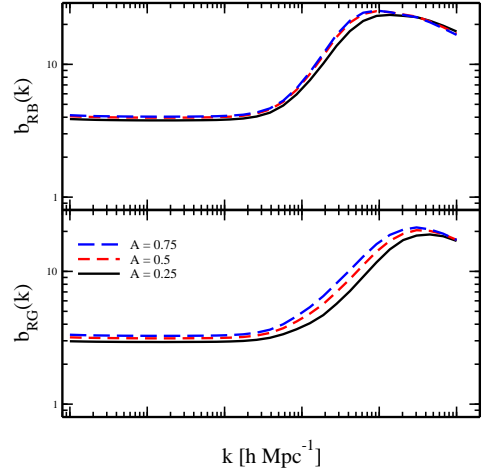}
\caption{$\brb$ and $\brg$ for several values of $A$.  $\Prr$ is unaffected
by our choice of $A$.}
\label{fig:bias_a}
\end{center}
\end{figure}

\begin{figure}
\begin{center}
\epsfxsize=2.45in \epsffile{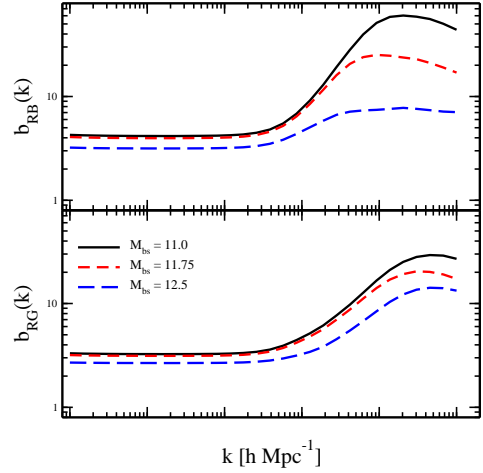}
\caption{$\brb$ and $\brg$ for several values of $\Mbs$.  Again, $\Prr$ remains
constant for different values of $\Mbs$.}
\label{fig:bias_mbs}
\end{center}
\end{figure}

Changing $A$ and $\Mbs$ has minimal effect on the red power spectrum, so 
long as we are in the $\Mr, \Mb \gg 10^{\Mbs}$ regime.  There are, however, 
effects on $\Pgg$ and $\Pbb$ (quite strong effects in the case of $\Mbs$) and 
we can begin to see our approximation neglecting the Gaussian term in 
Equation~\ref{eq:eta_prime_simple} break down as $\Mbs$ approaches $M_*$, 
resulting in biased $\Pbb$ relative to $\Pgg$ at small scales.  Increasing 
the value of $A$ has a slight global effect on $\brb$ and $\brg$ 
(Figure~\ref{fig:bias_a}), but mostly it controls the on-set of the break in 
the $\Pbb$ power-law with large values of $A$ lead to a sharper break.  The 
mass scale for the contribution of the Gaussian term in $\numgalb$ plays a 
much more significant role.  Large values of $\Mbs$ increase $\Pbb$ (and, to a 
lesser extent, $\Pgg$) on all scales, leading to a suppression of $\brb$ and 
$\brg$ as $\Mbs$ increases (Figure~\ref{fig:bias_mbs}).  Likewise, as the mass 
of the halos experiencing this boost in $\PPbb$ increases, the onset of the 
bump in the power law for $\Pbb$ occurs on larger and larger scales. 

As mentioned above, the over-riding theme of these variations appears to be 
the effect of changing at what scale and in what way the Poisson contribution 
to the power spectra sets in.  As $\Mb$ and $\Mr$ increase, fewer galaxies 
resulting from the power law parts of $\numgalb$ and $\numgalr$ populate the 
lower mass halos and small scale power-law behavior is suppressed in favor of 
the Gaussian contribution to $\numgalb$.  The degree to which this non-power 
law behavior in the red and blue galaxies is recreated in the data and 
simulations should be an excellent clue as to setting the relative amplitude 
of $\Mb$ and $\Mr$ as well as $\Mbs$.

\section{Conclusion}

We have shown that, through relatively simple modifications to the $\numgal$ 
and halo profile relations in the standard formalism, we can generate 
reasonable power spectra for red and blue galaxies, as well as a number 
of relative biases between the various galaxy power spectra.  By manipulation
of the parameters constituting the $\numgal$ relations, we have considerable
ability to modify the shapes of the red and blue power spectra.  Likewise, 
in the limit where we are not dominated by the $\numgal$ relations, our choice 
of halo profile parameters allows us to set the relative large-scale biases
between the various power spectra over a large range, while keeping the shapes
of the power spectra relatively constant.  There is some degree of degeneracy
between the choice of $\Mb$ \& $\Mr$ and the halo profiles for each of the 
galaxy sub-populations, suggesting that a simple fit to $\numgal$ from 
simulations without including this effect could result in an apparently 
larger difference between the two mass scales than the data actually indicate. 
Still, the degree of flexibility within the framework and sensitivity to the 
various input parameters suggest that measurements of these relative biases 
using current large galaxy survey data will provide strong constraints on 
these input parameters and the outputs of simulations.

\section{Acknowledgments}

Many thanks to Scott Dodelson and Ravi Sheth for useful advice and editing 
comments and thanks to Gilbert Holder for many insightful conversations.  
Additional thanks to Guinevere Kaufmann, Volker Springel and Antonaldo Diaferio
for assistance with the simulation data.

Support for this work was provided by the NSF through grant PHY-0079251 as 
well as by NASA through grant NAG 5-7092 and the DOE.

\section{References}

\def\refe {\par \hangindent=.7cm \hangafter=1 \noindent}
\def\apj { ApJ }
\def\astroph{{\tt astro-ph/}} 
\def\aap {A \& A }
\def\ajs{ ApJS }
\def\aj{AJ}
\def\prd{Phys ReV D}
\def\apjs{ ApJS }
\def\mnras { MNRAS }
\def\apjl { Ap. J. Let. }

\refe Benson, A.J., Cole, S., Frenk, C.S., Baugh, C.M., \& Lacey, C.G. 2000,
\mnras, 311, 793
\refe Bullock, J. S., Kolatt, T. S., Sigad, Y., Somerville, R. S., Kravtsov, 
A. V., Klypin, A. A., Primack, J. R., Dekel, A. 2001, \mnras, 321, 559
\refe Diaferio, A., Kauffmann, G., Colberg, J.M., \& White, S.D.M. 1999, 
\mnras, 307, 537
\refe Hamilton, A. J. S., Tegmark, M., Padmanabhan, N. 2000, \mnras, 317, 23
\refe Fukugita, M., Ichikawa, T., Gunn, J.E., Doi, M., Shimasaku, K., \& 
Scheider, D.P., 1996, \aj, 111,1748
\refe Kauffmann, G., Colberg, J.M., Diaferio, A. \& White, S.D.M. 1999, 
\mnras, 303, 188
\refe Jing, Y.P., Mo, H.J., Borner, G. 1998, \apj, 499,20
\refe Ma, C.-P. \& Fry, J.N., \apj, 543, 503
\refe Mo, H.J., White, S.D.M 1996, \mnras, 282, 1096
\refe Moore, B., Governato, F., Quinn, T., Stadel, J., \& Lake, G. 1999,
\mnras, 261, 827 
\refe Navarro, J., Frenk, C., \& White, S.D.M. 1996, \apj, 462, 563
\apjl 499, L5
\refe Peacock, J.A. \& Smith, R.E. 2000, \mnras, 318, 1144
\refe Press, W.H. \& Schechter, P. 1974, \apj, 187, 425
\refe Seljak, U. 2000, \mnras, 318, 203 
\refe Sheth, R., Diaferio, A., Hui, L., Scoccimarro, R. 2001, \mnras, 326, 463
\refe Sheth, R., Lemson, G. 1999, \mnras, 304, 767
\refe Sheth, R. \& Tormen, G. 1999, \mnras, 308, 119
\refe Scoccimarro, R., Sheth, R.K., Hui, L. \& Jain, B. 2001, \apj, 546, 20
\refe Somerville, R.S. \& Primack, J.R. 1999, \mnras, 310, 1087
\refe White, M., Hernquist, L., Springel, V. 2001, \apj, 550, 129
\refe White, S.D.M \& Frenk, C.S. 1991, \apj, 379, 52
\refe White, S.D.M \& Rees, M.J. 1978, \mnras, 183, 341

\end{document}